\documentclass[A4paper, 12pt]{article}
\usepackage[margin=1in]{geometry}
\usepackage[numbers,sort&compress]{natbib} 
\usepackage{authblk} 
\usepackage{color}
\usepackage{amsmath}
\usepackage{amssymb} 
\usepackage{graphicx} 
\usepackage{subfigure}
\usepackage{fancyhdr} 
\usepackage{amsfonts}
\usepackage{tabularx} 
\usepackage{slashed}
\usepackage[
singlelinecheck=false 
]{caption}

\usepackage[utf8]{inputenc}

\begin{document}
\title{\huge \bf{
Holographic Integral Geometry \mbox{with Time Dependence} 
}}
\author{Bart{\l}omiej Czech$^1$\footnote{bartlomiej.czech@gmail.com}}
\author{ Yaithd D. Olivas$^2$\footnote{dolivas@live.com.mx}}
\author{Zi-zhi Wang$^1$\footnote{wang-zz17@mails.tsinghua.edu.cn}}
\affil{$^1$\emph{Institute for Advanced Study, Tsinghua University}\\\emph{Beijing 100084, China}}
\affil{$^2$\emph{Departamento de F\'{i}sica de Altas Energ\'{i}as, Instituto de Ciencias Nucleares, }\\\emph{Universidad Nacional Aut\'{o}noma de M\'{e}xico}\\\emph{Apartado Postal 70-543, CDMX 04510, M\'{e}xico}}

\maketitle
\thispagestyle{fancy}
\pagenumbering{gobble} 

\begin{abstract}
\noindent \parskip 5pt
We write down Crofton formulas---expressions that compute lengths of spacelike curves in asymptotically AdS$_3$ geometries as integrals over kinematic space---which apply when the curve and/or the background spacetime is time-dependent. Relative to their static predecessor, the time-dependent Crofton formulas display several new features, whose origin is the local null rotation symmetry of the bulk geometry. In pure AdS$_3$ where null rotations are global symmetries, the Crofton formulas simplify and become integrals over the null planes, which intersect the bulk curve.  
\end{abstract}

\newpage
\pagenumbering{arabic} 

\section{Introduction}
\label{Intro}

Recent years have taught us much about the emergence of space, but not nearly as much about the nature of time. In holographic duality, key insights concerning the microscopic fabric of spacetime---the Ryu-Takayanagi proposal \cite{rt, rt2} with its various reformulations and generalizations \cite{hrt, minimax, bitthreads}, holographic entropy inequalities \cite{monogamy, hec}, error correction \cite{errorcorr} or the analogy to tensor networks \cite{briantns, errortn, pluperfect}---all stipulate the existence of some preferred spatial slice of the bulk geometry. A true understanding of the microscopic underpinning of gravity should give us an equally deep and detailed perspective on the timelike direction---say, the lapse and shift in the ADM language. Some works have begun to address this issue \cite{searchforbulkpt, nullcuts, covtn, modflow}, but we are still far away from answering the question: If entanglement manifests itself as space \cite{marksessay, erepr}, what concept manifests itself as time? 

Barring some unanticipated breakthrough, a natural route toward answering this question is to find covariant versions of statements, which are currently only known to hold on spatial slices of gravitational spacetimes. This paper takes on one such statement: the Crofton formula \cite{intgeom}. It says that the length of a spacelike curve on a static slice of an asymptotically AdS$_3$ geometry `counts' the geodesics which intersect the said curve. In holographic theories, the correct measure for this `counting' problem turns out to have a direct information theoretic meaning on the boundary: it is the conditional mutual information of regions, which are selected by the geodesics. This is a powerful lesson about the information theoretic origin of the notion of distance in the bulk \cite{nutsandbolts, infolength}, which has led to a number of interesting insights and follow-ups \cite{deBoer:2015kda, Czech:2015xna, tnks, neleandclaire, opeblocks, opeblocks2, boundaryks, Czech:2017ryf, Cresswell:2017mbk, alberto, johanna1, mbc, Abt:2018ywl, entholonomies, nele, Cresswell:2018mpj, claire, sewingkit}. However, the scope of this lesson has been mostly limited to the static setup.\footnote{Refs.~\cite{opeblocks, opeblocks2, alberto} have applied kinematic space (space of geodesics) techniques in time-dependent settings, but they exploited an integrated version of the Crofton formula---differential entropy \cite{differentialentropy, hholes}---without writing down the Crofton formula explicitly.}

In this paper we write down Crofton formulas for a spacelike curve in AdS$_3$, which do not assume that the curve lives on a static slice of the bulk geometry. These covariant Crofton formulas have a lot of  interesting features, one of which is that there are many such formulas for a single curve! Different formulas that compute the length of the same curve are related to one another by a certain `gauge freedom,' which is generated in the bulk by local null rotations.\footnote{This `gauge freedom' should not be confused with the modular gauge symmetry and the associated modular Berry connection, whose holonomies are computed by eq.~(\ref{diffent}); see \cite{mbc, sewingkit}.} We will decode this statement at various stages of the text, starting with Sec.~\ref{NVACondition}.

A second interesting fact about the covariant Crofton formulas is that we do not integrate over the geodesics that intersect the curve. In a generic asymptotically AdS$_3$ geometry one can take many different regions of integration and none of them favors geodesics that intersect the curve. We have not found a unifying geometric characterization of all admissible integration regions except in pure AdS$_3$, where the Crofton formula---instead of integrating over intersecting geodesics---integrates over all null planes (homogeneous lightsheets) that intersect the curve. As we explain below, these facts too originate from the null rotation symmetry of the bulk geometry.

The paper is organized as follows: Section~\ref{ReviewNVAC} reviews the necessary background material---differential entropy, the static Crofton formula, null rotations and kinematic space. In Section~\ref{sec:covariant} we write down the covariant Crofton formulas for general horizonless, asymptotically AdS$_3$ geometries. Section~\ref{ads3sec} explains the simplifications that occur in pure AdS$_3$, with the final result that the length of a spacelike curve in pure AdS$_3$ `counts' the null planes that intersect the curve. We close with a Discussion.

\section{Review}
\label{ReviewNVAC}

The setup of this paper is the AdS$_3$/CFT$_2$ correspondence. We assume that the low energy bulk theory is Einstein gravity so that entanglement entropies of CFT intervals are computed by lengths of bulk geodesics \cite{rt, rt2, hrt, minimax}. The starting point is the differential entropy formula \cite{differentialentropy, hholes}, which expresses the length of a general spacelike bulk curve in terms of lengths of geodesics or, by the Ryu-Takayanagi proposal, in terms of entanglement entropies of CFT intervals. 

In this section we review differential entropy as well as other concepts, which will be useful in the remainder of the paper. The presentation in Sec.~\ref{NVACondition} is partly new and complements the findings of Ref.~\cite{hholes}.

\subsection{Differential Entropy}
\label{sec:diffent}

Consider a smooth, closed spacelike curve in the bulk of an asymptotically AdS$_3$, horizonless geometry. For convenience, we will also assume a certain notion of convexity to be defined momentarily.  By smoothness, every point on the curve has a tangent geodesic; we denote the boundary coordinates of its endpoints with:
\begin{equation}
y_L(\lambda) = (z_L(\lambda), \bar{z}_L(\lambda)) 
\qquad {\rm and} \qquad 
y_R(\lambda) = (z_R(\lambda), \bar{z}_R(\lambda))
\label{defyy}
\end{equation}
Here $\lambda$ is a parameter around the curve and $z, \bar{z}$ are lightlike coordinates on the boundary cylinder: 
\begin{equation}
z = \theta + t
\qquad {\rm and} \qquad 
\bar{z} = \theta - t\,.
\label{lightcoords}
\end{equation} 
The subscripts $L$ and $R$ mark the left and right endpoints of the geodesic, as seen from the boundary interval $(y_L(\lambda), y_R(\lambda))$ looking into the bulk. Throughout this paper we will consider only oriented geodesics, so that we can unambiguously say that geodesic~(\ref{defyy}) subtends the CFT interval $(y_L(\lambda), y_R(\lambda))$ and not $(y_R(\lambda), y_L(\lambda))$. The family of oriented geodesics~(\ref{defyy}) (equivalently, the family of subtended intervals $(y_L(\lambda), y_R(\lambda))$) is the one whose entanglement wedges meet the curve at exactly one point each. (With the other orientation, the entanglement wedges of the subtended intervals would each have contained the entire bulk curve.) In the 2+1-dimensional, time-dependent context, the convexity of the curve will mean for us that this condition can be globally satisfied. 

Under these assumptions, the length of the curve equals \cite{differentialentropy, hholes}:
\begin{equation}
{\rm length} = \int d\lambda\, \frac{dy_R^\mu}{d\lambda}\, \frac{\partial S(y_L(\lambda), y_R)}{\partial y_R^\mu} \Big|_{y_R = y_R(\lambda)}\,,
\label{diffent}
\end{equation}
where the summed index $\mu = 0,1$ is shorthand for $y^0 = z$ and $y^1 = \bar{z}$. Quantity $S(y_L, y_R)$ is the length of the bulk geodesic that connects $y_L$ and $y_R$ on the boundary. When the geodesic is minimal and the homology constraint is satisfied, this is equal to the entanglement entropy of the CFT interval $(y_L, y_R)$. We set $4G_N \equiv 1$ throughout. 

\begin{figure}[t]
\centering
\includegraphics[width=.4\textwidth]{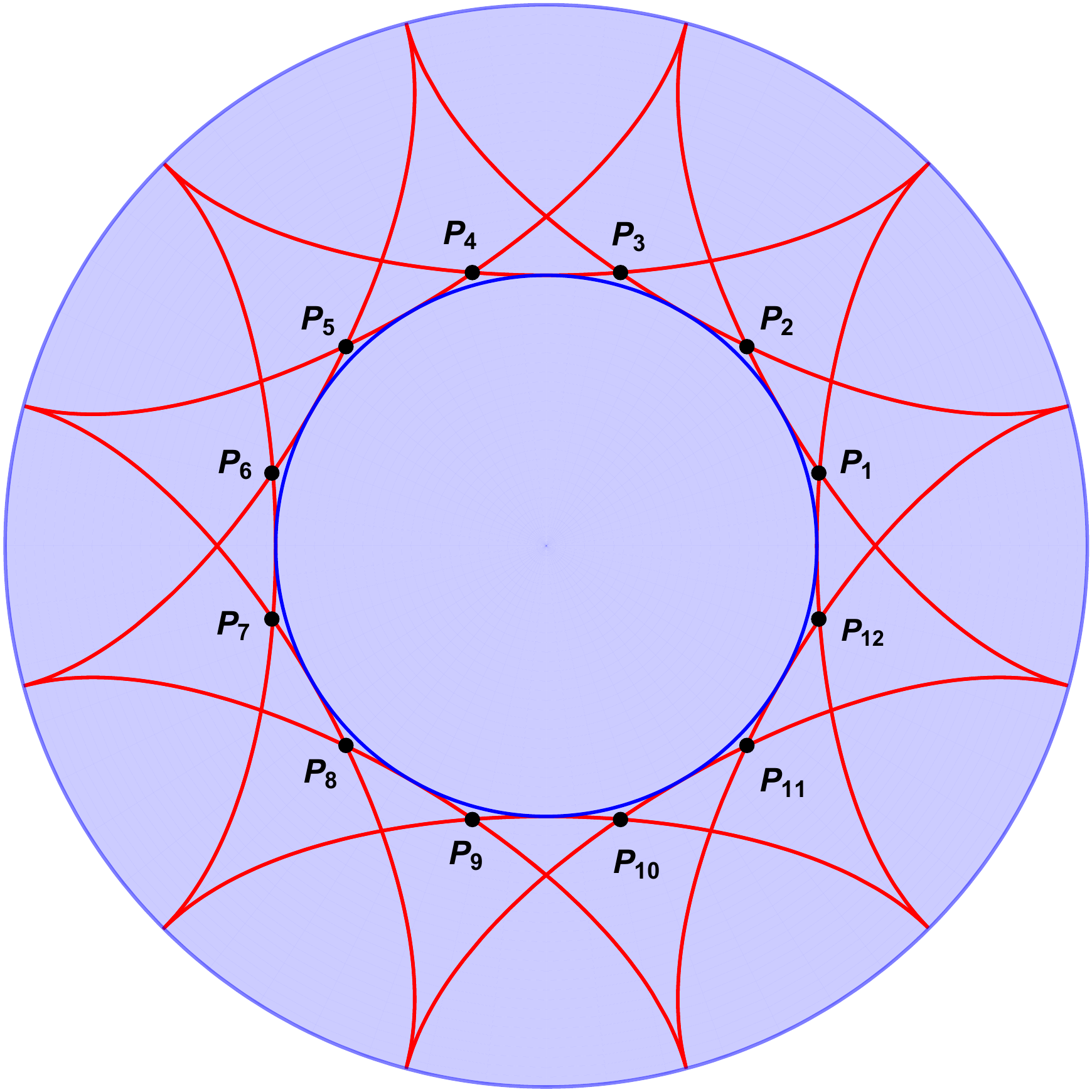}
\hfill
\includegraphics[width=.4\textwidth]{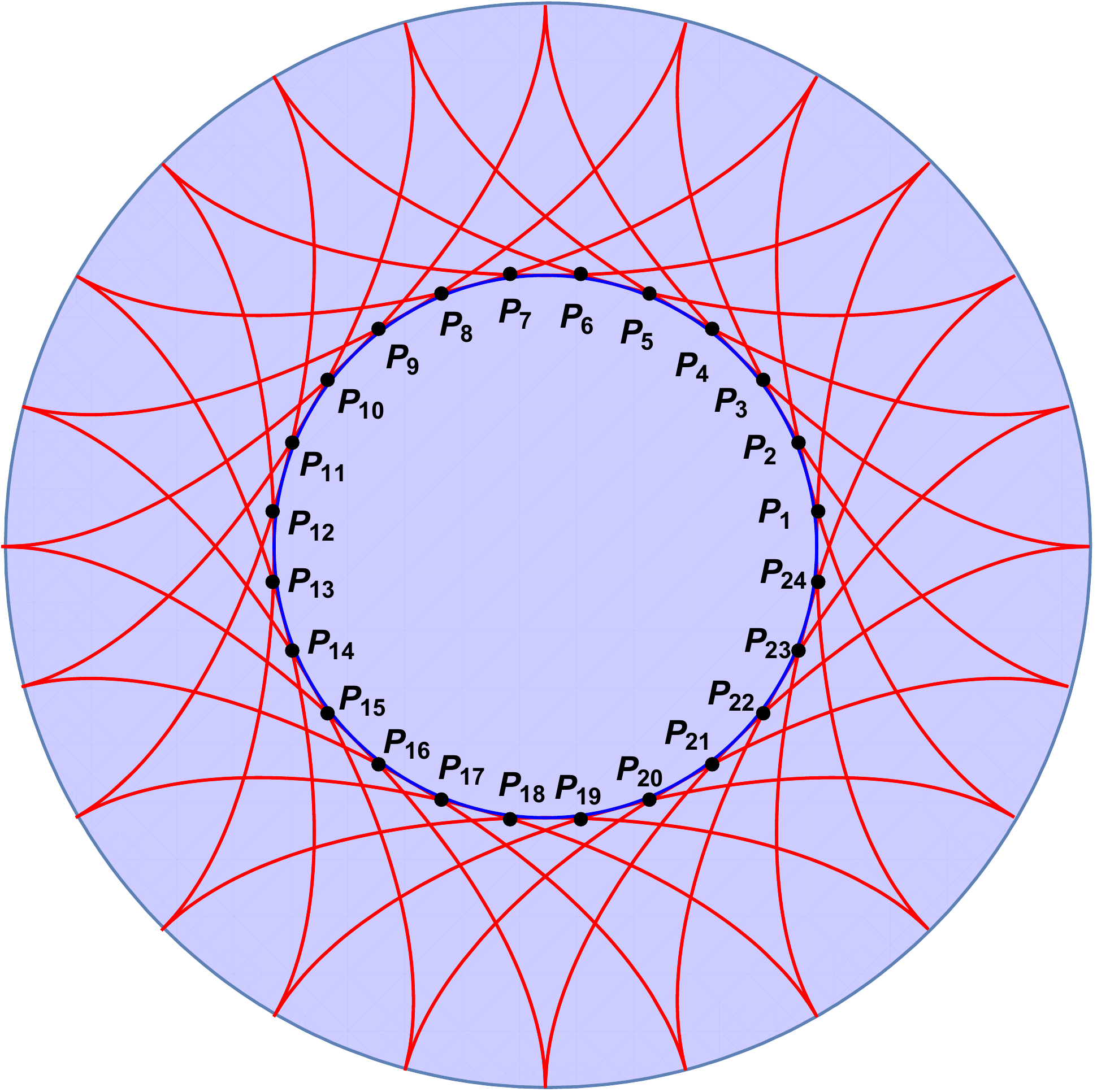}
\caption{Two sequences of geodesics tangent to a common bulk static curve. On the right, the sequence is finer and the intersection points of consecutive pairs of geodesics live closer to the bulk curve. In the continuum, the intersection points approach the bulk curve, which is identified with the common envelope of the geodesics. }
\label{FinerSequence}
\end{figure}

Eq.~(\ref{diffent}) is the differential entropy formula. It is useful to inspect briefly the geometry underlying it. First, suppose the bulk geometry is static and consider a bulk curve contained in a static slice. 
In this case, $dy_R / d\lambda$ points in the spacelike ($\theta$) boundary direction and the formula simplifies upon setting the arbitrary parameter $\lambda \equiv \theta_R$:
\begin{equation}
{\rm length} = \int d\theta_R \frac{\partial S(\theta_L, \theta_R)}{\partial \theta_R} \Big|_{\theta_L = \theta_L(\theta_R)}
\label{diffentstatic}
\end{equation}
The function $\theta_L(\theta_R)$ simply picks the geodesics tangent to the bulk curve. Now consider a discrete subset of them, with coordinates $\theta_L = \theta_L (\theta_R^{(i)})$ and $\theta_R = \theta_R^{(i)}$. The consecutive geodesics intersect on the bulk static slice; call the intersection of the $(i-1)^{\rm th}$ and $i^{\rm th}$ geodesic $P_i$. In the limit in which the spacing between consecutive geodesics becomes finer, the points $P_i$ approach the bulk curve and the integrand of (\ref{diffentstatic}) becomes, up to a total derivative, the infinitesimal distance between $P_i$ and $P_{i+1}$---that is, the length element along the curve. This is illustrated in Fig.~\ref{FinerSequence}. 

When the curve does not live on a static slice---or when the background geometry is not static---the geometric picture in Fig.~\ref{FinerSequence} must be modified. Indeed, if the curve is not confined to a static two-dimensional submanifold of the bulk, the consecutive geodesics will not in general intersect. We explain the requisite modification of Fig.~\ref{FinerSequence} after introducing one further generalization of formula (\ref{diffent}).

\subsection{Null Vector Alignment}
\label{NVACondition}

This subsection reviews and extends the material of \cite{hholes}.  That reference showed that the points $y_L(\lambda)$ and $y_R(\lambda)$ can be chosen in other ways---their connecting geodesic not tangent to the bulk curve---and still satisfy eq.~(\ref{diffent}). The condition to be imposed, which generalizes tangency, is called `null vector alignment' (NVA).

\begin{figure}[t]
\centering
\includegraphics[width=.50\textwidth]{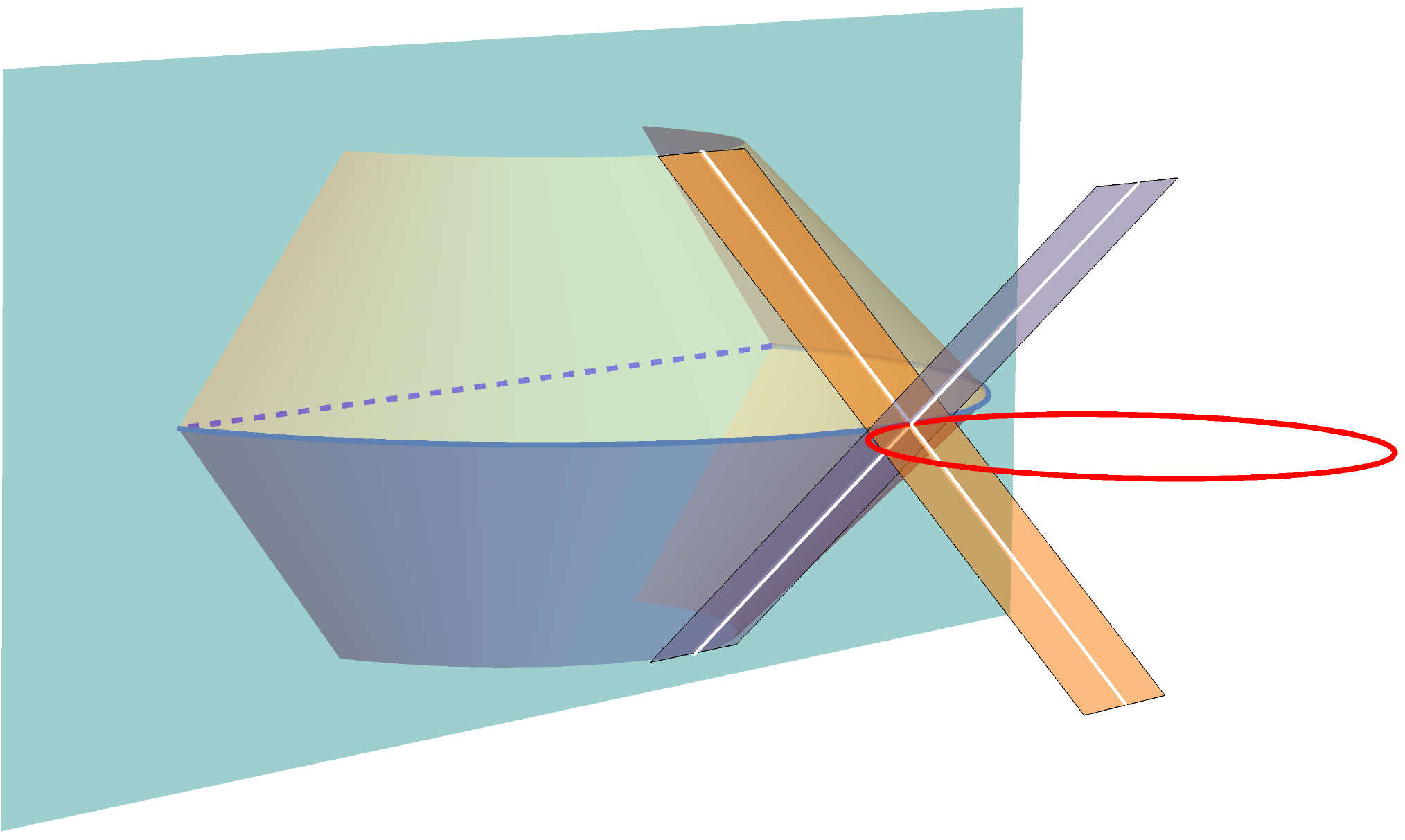}
\hfill
\includegraphics[width=.42\textwidth]{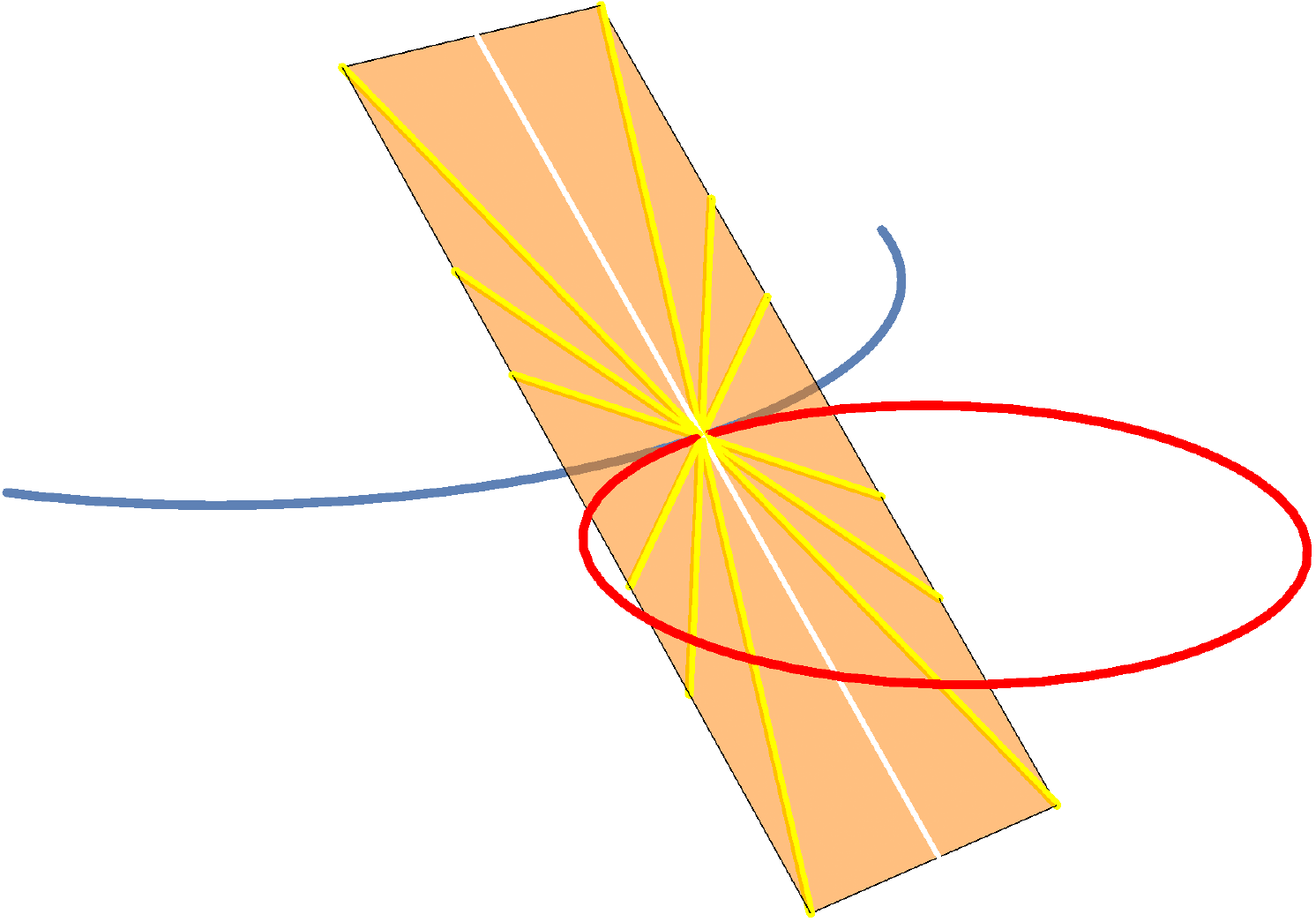}
\caption{A bulk curve (red) has two orthogonal lightsheets emanating from it. In the left panel, we display narrow strips of the outgoing orthogonal lightsheet (orange) and the ingoing orthogonal lightsheet (gray) in a neighborhood of a point $\lambda$, highlighting the orthogonal null rays generated by vectors $n_o$ and $n_i$ (white lines). We also show the lightsheets emanating from the tangent geodesic (blue). In the right panel, we show the NVA geodesics tangent to the outgoing lightsheet at $\lambda$. These geodesics are locally related to one another by a null rotation, which fixes the outgoing orthogonal null vector $n_o$. In the limit of infinite null rapidity, the NVA geodesics approach the outgoing orthogonal null vector $n_o$. The unique geodesic which is NVA with respect to both lightsheets is the tangent (blue) geodesic.} 
\label{NullSheetsNVAGeodesics}
\end{figure}

Null vector alignment at point $\lambda$ on the curve means that the geodesic passes through $\lambda$ and that it is tangent \emph{to the lightsheet emanating from the curve.} Equivalently, null vector alignment can be stated as the tangency of two lightsheets---one emanating from the curve and one from the geodesic; see Fig.~\ref{NullSheetsNVAGeodesics}.
This type of relation between the curve and a geodesic is an inherently Lorentzian concept; its only Euclidean analogue is if the curve and the geodesic are tangent to one another. 

\paragraph{Two families of NVA geodesics} 
Note that the curve has two lightsheets emanating from it. Assuming that the curve is closed and convex, we can label one of them the outgoing lightsheet and the other the ingoing lightsheet; see Fig.~\ref{NullSheetsNVAGeodesics}. Consequently, at any given point on the curve there are two families of NVA geodesics: one family tangent to the outgoing lightsheet and one family tangent to the ingoing lightsheet. There is one geodesic which is common to both families: because it follows the intersection of both lightsheets, it is the geodesic tangent to the curve. 

\paragraph{Null rotations} All geodesics that are null vector-aligned (NVA) at $\lambda$ are related to the geodesic tangent at $\lambda$ by a transformation, which is locally a null rotation. To understand this fact in more detail, refer to Fig.~\ref{NullSheetsNVAGeodesics} and consider a neighborhood of the point $\lambda$ small enough to be treated as flat space, so nomenclature from the 2+1-dimensional Lorentz group will apply. At $\lambda$, our curve selects a privileged triple of vectors (a triad): the curve's tangent $t$ and two null vectors  $n_o$ and $n_i$ orthogonal to the curve. Locally, the outgoing lightsheet is a null plane generated by the tangent vector $t$ and by $n_o$. The normal vector to this null plane is $n_o$ itself; because it is null, $n_o$ both lives on the null plane and is normal to it. The family of NVA geodesics that are tangent to the outgoing lightsheet therefore have one thing in common: they are orthogonal to the null vector $n_o$. The Lorentz transformation, which locally relates to one another this family of NVA geodesics, must therefore preserve the vector $n_o$. Of course, the same analysis applies to the other family of NVA geodesics, with the replacement $n_o \to n_i$. 

In 2+1 dimensions, a rotation fixes a timelike vector while a boost fixes a spacelike vector. Transformations that fix a null vector are a distinct conjugacy class of the Lorentz group called `null rotations.' Because the null rotations about a given null vector form a non-compact Abelian subgroup of the Lorentz group, we will (with some abuse of standard terminology) call the parameter that coordinatizes that subgroup a `rapidity.' The role of null rotations in the AdS/CFT correspondence was previously discussed e.g. in \cite{joannullrot}.

In case null rotations seem unfamiliar, we illustrate them with the following example. Consider a small neighborhood of a point $\lambda$ on the bulk curve; we assume the neighborhood small enough to be treated as flat. With an appropriate choice of coordinates, the vectors discussed in the previous paragraphs can be written as
\begin{equation}
n_o = \left( \begin{array}{r} 1 \\ 1 \\ 0 \end{array} \right)
\qquad {\rm and} \qquad
n_i = \left( \begin{array}{r} 1 \\ -1 \\ 0 \end{array} \right)
\qquad {\rm and} \qquad
t = \left( \begin{array}{r} 0 \\ 0 \\ 1 \end{array} \right),
\end{equation}
where we take the metric to be ${\rm diag}(-1,1,1)$. An $SO(1,2)$ transformation that fixes $n_o$ and maps different NVA geodesics from the outgoing family to one another can be written as:
\begin{equation}
N = \left( \begin{array}{ccr} 1+\rho^2/2 & -\rho^2/2 & -\rho \\ \rho^2/2 & 1-\rho^2/2 & -\rho \\ -\rho & \rho & 1 \end{array} \right),
\end{equation}
where $\rho$ parameterizes the `null rapidity' and ranges from $-\infty$ to $+\infty$. Explicitly, we have:
\begin{equation}
N n_o = n_o \qquad {\rm and} \qquad N n_i = n_i + \rho^2 n_o - 2\rho\, t \qquad {\rm and} \qquad
N t = t - \rho\, n_o  
\end{equation}
This last equation is the NVA condition, stated in the same language as eq.~(4.14) in \cite{hholes}. In there, the authors described the NVA condition as the demand that the normalized tangent vector to the bulk curve (vector $t$) and the tangent to the NVA geodesic (vector $N t$) differ only by a multiple of an orthogonal null vector (here $n_o$). 

We should remember, however, that a null rotation maps different NVA geodesics to one another only in a small neighborhood of the point $\lambda$. One exception is pure AdS$_3$, in which any null geodesic is related to any other by a global isometry. This feature, which we exploit extensively in Section~\ref{ads3sec}, will allow us to make stronger statements in locally AdS$_3$ spacetimes.  

\paragraph{From geodesics to curves} Given a family of geodesics with endpoints $y_L(\lambda)$ and $y_R(\lambda)$, what is the curve whose length eq.~(\ref{diffent}) computes? Equivalently, how to find a curve which is NVA to a given continuous family of geodesics? Ref.~\cite{hholes} answered this question by generalizing the static construction reviewed in Sec.~\ref{sec:diffent}, which involved a sequence $P_i$ of intersection points of consecutive tangent geodesics. The argument there left out curves with time dependence (even when tangent geodesics and not NVA geodesics are used); the construction of \cite{hholes} also covers this special case. 

\begin{figure}
\centering
\includegraphics[width=.42\textwidth]{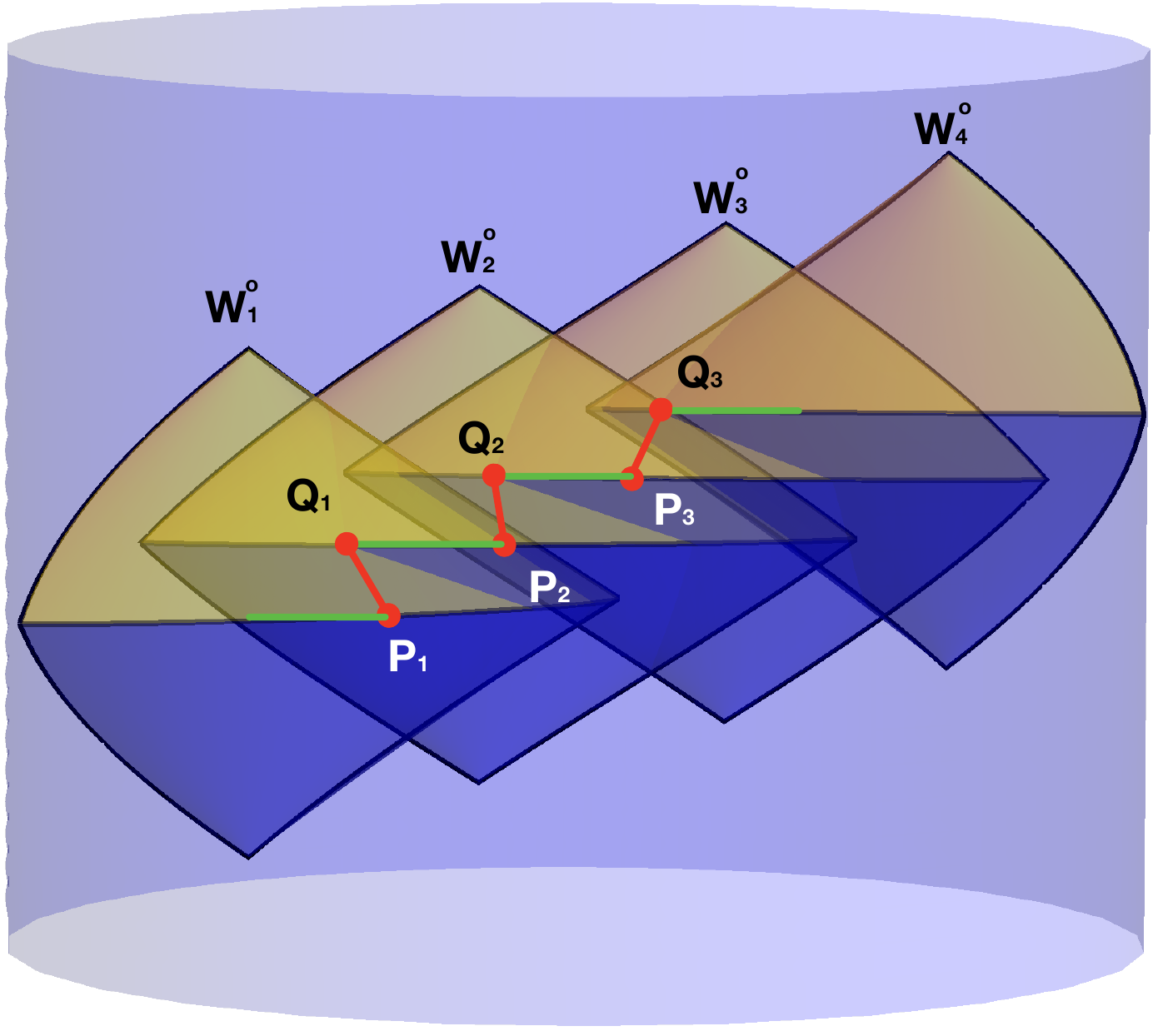}
\hfill
\includegraphics[width=.42\textwidth]{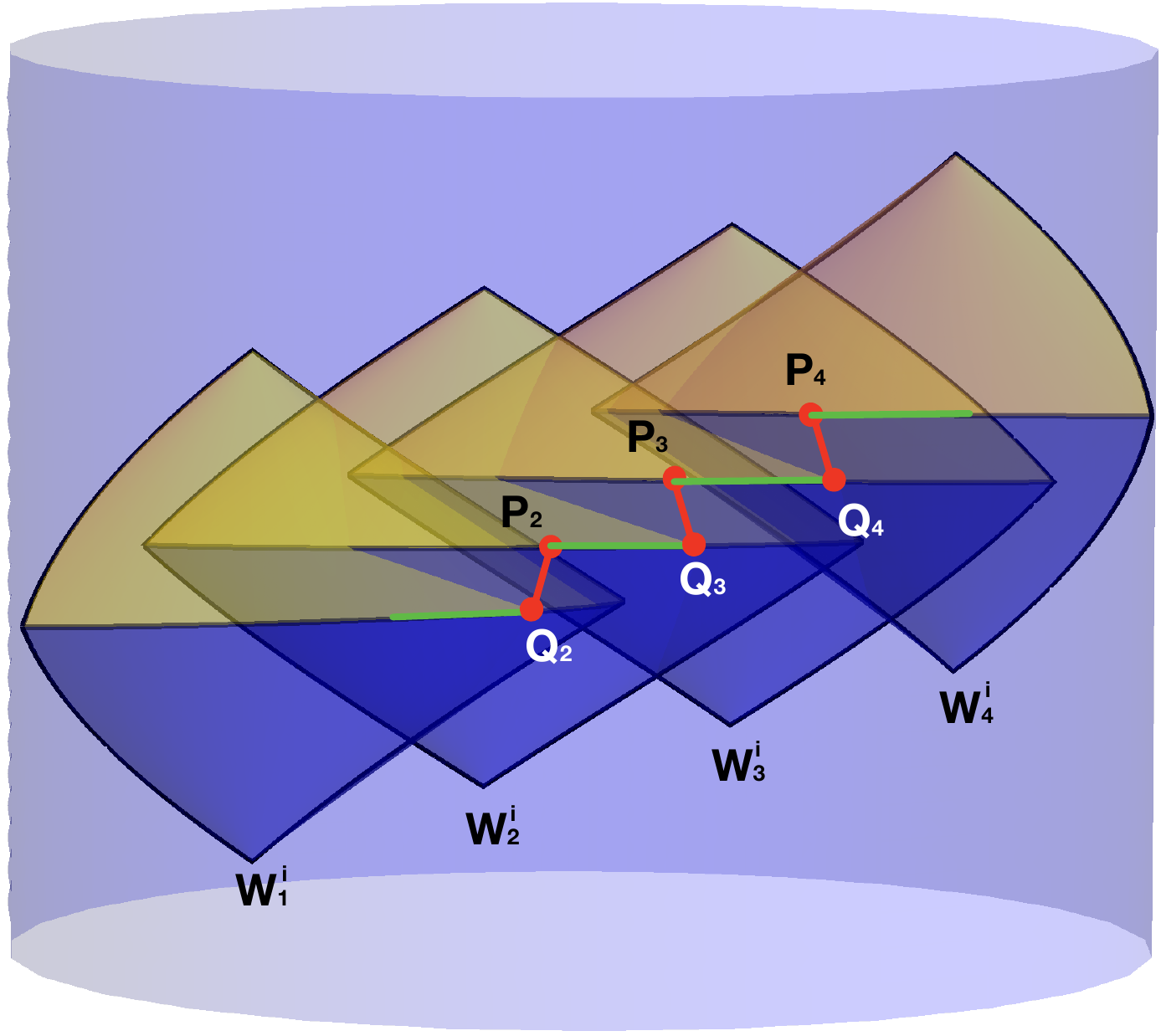}
\caption{How the same sequence of geodesics defines a bulk curve using outgoing (left) and ingoing (right) lightsheets in the NVA condition. The green lines are pieces of geodesics $\lambda$ and the red lines are null segments that join these pieces. In the continuum limit, the pieces of geodesics and lightrays become the bulk curve. The constructions with outgoing and ingoing lightsheets typically generate different bulk curves. }
\label{NVAConstruction} 
\end{figure}

For illustration, refer to Fig.~\ref{NVAConstruction}. To each geodesic (labeled by $\lambda$) assign a lightsheet emanating from it; call it $W(\lambda)$. As we emphasized before, every geodesic has two such lightsheets; the choice of $W(\lambda)$ should be continuous. As in Sec.~\ref{sec:diffent}, consider a discrete progression of geodesics and lightsheets labeled by a sequence $\lambda^{(i)}$. The $\lambda^{(i-1)}$-geodesic meets the $W(\lambda^{(i)})$ lightsheet at a point, which we call $Q_i$. Now follow the unique lightray, which is contained in $W(\lambda^{(i)})$ and passes through $Q_i$, until it meets the geodesic $\lambda^{(i)}$; call that meeting point $P_i$. In the limit of increasingly fine spacing between consecutive $\lambda^{(i)}$s, the sequence of points $P_i$ will converge to a continuous curve that satisfies eq.~(\ref{diffent}).

We offer a few comments on this construction: In the static discussion in Sec.~\ref{sec:diffent}, the points $Q_i$ and $P_i$ coincided. When we exploit the NVA freedom or describe a time-dependent space-like curve using tangent geodesics, the points $Q_i$ and $P_i$ do not coincide except in the continuum limit. Finally and most importantly, a given sequence of geodesics generally picks out two distinct bulk curves to which they are NVA: one constructed by taking the $W(\lambda)$s to be the outgoing lightsheets and one from the ingoing lightsheets. 

\paragraph{From curves to geodesics} 
For a given bulk curve, there is a large freedom in choosing a continuous family of NVA geodesics. Each such choice of $y_L(\lambda)$ and $y_R(\lambda)$ will of course satisfy eq.~(\ref{diffent}). The first freedom is discrete: the geodesics can be taken from the outgoing or the ingoing lightsheets. The further freedom is in choosing a continuous, $\lambda$-dependent rapidity parameter, which sets the magnitude of the local null rotation separating the NVA geodesics from the tangent geodesics at each $\lambda$. 

\subsection{The Crofton Formula}
\label{CroftonTimeSlice}

The differential entropy formula recasts the length of a closed bulk curve as a one-dimensional integral over a sequence of NVA geodesics. When the curve is static and tangent (not general NVA) geodesics are used, it can be converted to a two-dimensional integral over geodesics that intersect the curve; this is the Crofton formula \cite{intgeom}. Our interest in this paper is in generalizing this picture to setups with time dependence. 

What converts a one-dimensional integral into a two-dimensional integral is Stokes's theorem. One way to apply it to the static version of the differential entropy formula~(\ref{diffentstatic}) is to write
\begin{equation}
{\rm length} = \int_{?}^{\theta_L = \theta_L(\theta_R)}\, d\theta_R d\theta_L\, 
\frac{\partial^2 S(\theta_L, \theta_R)}{\partial \theta_R\, \partial \theta_L} 
\label{crofton1}
\end{equation}
and choose the lower limit of integration marked `?' so that
\begin{equation}
\int d\theta_R \frac{\partial S(\theta_L, \theta_R)}{\partial \theta_R} \Big|_{?} = 0\,.
\end{equation}
This is tantamount to choosing a second `curve' of zero length and subtracting it from equation~(\ref{diffentstatic}). More generally, any expression of the form~(\ref{crofton1}) computes the difference between lengths of two curves: the one defined by the upper and the lower limit of integration:
\begin{equation}
{\rm length}^{\rm upper} - {\rm length}^{\rm lower} 
= \int^{\theta_L = \theta_L^{\rm upper}(\theta_R)}_{\theta_L = \theta_L^{\rm lower}(\theta_R)}\, d\theta_R d\theta_L\, 
\frac{\partial^2 S(\theta_L, \theta_R)}{\partial \theta_R\, \partial \theta_L} 
\label{croftongen}
\end{equation}
Applying Stokes's theorem to rewrite formula (\ref{diffentstatic}) necessarily requires a choice of a second limit of integration because the loop $\{(\theta_L(\theta_R), \theta_R)\}_{\theta_R}$ is not contractible in the space of geodesics on a static slice of AdS$_3$ (kinematic space). If it were contractible---if a homotopy from $\{(\theta_L(\theta_R), \theta_R)\}_{\theta_R}$ to the trivial loop could be found---then the projection $(\theta_L, \theta_R) \to \theta_R$ would produce a homotopy, which contracts a loop around a circle to a point.

\begin{figure}[t]
\centering
\includegraphics[width=.45\textwidth]{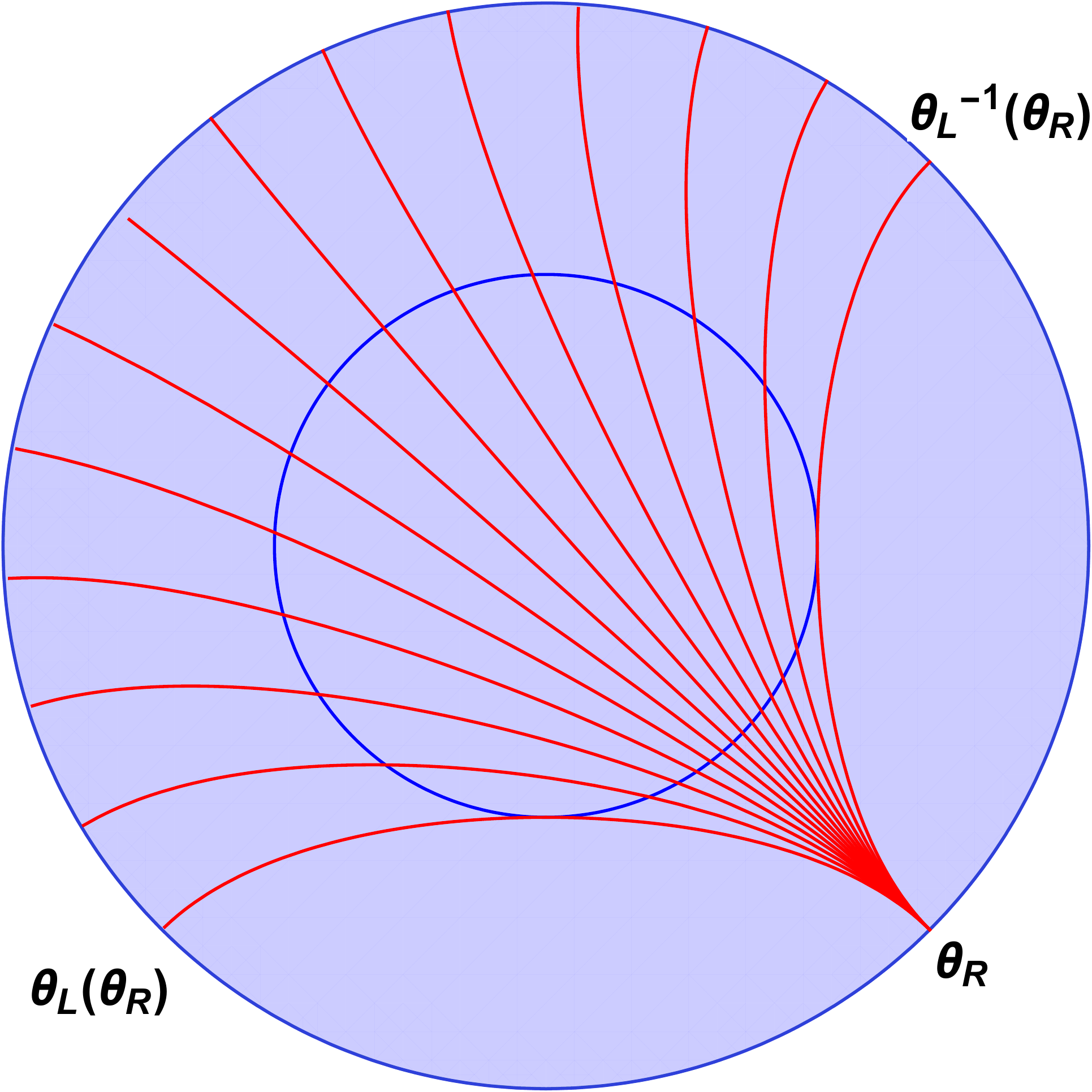}
\caption{In eq.~(\ref{croftongen}), setting the upper limit of integration to the locus $\theta_L = \theta_L(\theta_R)$ and the lower limit to $\theta_L = \theta_L^{-1}(\theta_R)$ results in an integration region, which encompasses all oriented geodesics that intersect the bulk curve. Here we display which geodesics are integrated over for a fixed value of $\theta_R$.}
\label{StaticCrofton}
\end{figure}

In a horizon-free geometry (dual to a CFT pure state),\footnote{In the presence of a horizon, the RT geodesic with endpoints $(y_L, y_R) = (y_1, y_2)$ is different from the one with endpoints $(y_L, y_R) = (y_2, y_1)$.} the most natural way of using (\ref{croftongen}) to compute the length of a given bulk curve is to let both limits of integration sweep the tangent geodesics, but one with endpoints reversed. Explicitly, if the upper limit of integration is the locus $\theta_L = \theta_L(\theta_R)$ then the lower limit of integration sets $\theta_L = \theta_L^{-1}(\theta_R)$. On the latter integration contour, eq.~(\ref{diffentstatic}) evaluates to the length of the original bulk curve with an extra minus sign for the reversal of orientation. The resulting integration region in eq.~(\ref{croftongen}) encompasses all the geodesics on the static slice that intersect the bulk curve; see Fig.~\ref{StaticCrofton}. 

In sum, we arrive at the following \emph{Crofton formula} for the length of a closed spacelike curve living on a static slice of the bulk geometry:
\begin{equation}
{\rm length} 
= \frac{1}{2} \int_\textrm{\{geodesics that intersect the bulk curve\}} \, d\theta_R d\theta_L\, 
\frac{\partial^2 S(\theta_L, \theta_R)}{\partial \theta_R\, \partial \theta_L} 
\label{crofton}
\end{equation}
We shall soon write down an analogue of this formula for curves, which do not live on a static slice of the bulk. 

\paragraph{Comment} The integrand in eqs.~(\ref{crofton1}), (\ref{croftongen}) and (\ref{crofton}) has a direct interpretation in quantum information theory. It is the conditional mutual information of two infinitesimal intervals of length $d\theta_L$ and $d\theta_R$, conditioned on the interval $(\theta_L, \theta_R)$. This object is guaranteed to be positive by the strong subadditivity inequality \cite{ssaproof, ssahologr}. We may interpret eq.~(\ref{crofton}) as a `count' of geodesics that intersect the bulk curve, with a measure supplied by quantum information theory.

\subsection{Kinematic Space}
\label{4dksintro}
The static Crofton formula (\ref{crofton}) sweeps geodesics drawn from a single slice of the bulk geometry. In seeking a generalization to the time-dependent case, we will have to go outside these restricted settings and consider the space of \emph{all} oriented spacelike geodesics. This general, four-dimensional kinematic space was studied in \cite{opeblocks, opeblocks2}. 

Its coordinates are $z_L, \bar{z}_L, z_R, \bar{z}_R$, which we defined in Sec.~\ref{sec:diffent}. In applying Stokes' theorem to the differential entropy formula (\ref{diffent}), we will encounter the 2-form, which is the exterior derivative of its integrand:
\begin{equation}
\omega = (\partial_L \partial_R S) dz_L dz_R + (\bar{\partial}_L \bar{\partial}_R S) d\bar{z}_L d\bar{z}_R 
+  (\bar{\partial}_L \partial_R S) d\bar{z}_L dz_R + (\partial_L \bar{\partial}_R S) dz_L d\bar{z}_R\,,
\label{4dcroftonform}
\end{equation}
where $\partial_L = \partial_{z_L}$ and $\bar{\partial}_L = \partial_{\bar{z}_L}$ and likewise for $\cdot_{R}$. 

A big simplification occurs in the case of the CFT$_2$ ground state and its Virasoro descendants: the entanglement entropy decomposes into two pieces, which depend only on $z_L, z_R$ (respectively $\bar{z}_L, \bar{z}_R$). As a result, the two last terms in (\ref{4dcroftonform}) drop out. Since $\omega$ has no joint $z, \bar{z}$-dependence, we can represent the kinematic space of the vacuum and its descendants as a product of two topological cylinders, one coordinatized by $z_L$ and $z_R$ and the other by $\bar{z}_L$ and $\bar{z}_R$. The geometry of this factorized kinematic space was discussed in detail in \cite{opeblocks}. 

\section{Covariant Crofton Formula}
\label{sec:covariant}

In the previous section, we went from the static differential entropy equation~(\ref{diffentstatic}) to the Crofton integral~(\ref{crofton}) using Stokes's theorem. We will do the same to find the covariant Crofton formula. 

The integrand will be the form $\omega$ we encountered in eq.~(\ref{4dcroftonform}), i.e. the exterior derivative of the covariant differential entropy integrand from eq.~(\ref{diffent}). To enclose a compact two-dimensional region of integration, we again need two boundary contours.\footnote{The reason, as in the static case, is topological. If we could contract the loop $\{(y_L^o(\lambda),\, y_R^o(\lambda))\}_\lambda$ then the projection $(y_L, y_R) \to y_R$ would produce a homotopy, which contracts a loop wrapping around the boundary cylinder to a single boundary point.} We will represent them as two parametric curves in kinematic space:  $(y_L^o(\lambda), y_R^o(\lambda))$ and $(y_L^{\tilde{i}}(\lambda), y_R^{\tilde{i}}(\lambda))$. Each of them individually, when plugged into eq.~(\ref{diffent}), computes the length of some bulk curve. Thus, Stokes's theorem tells us that: 
\begin{equation}
{\rm length}^{(o)} - {\textrm{length}}^{(\,\tilde{i}\,)} 
= \int^{(y_L^o(\lambda),\, y_R^o(\lambda))}_{(y_L^{\tilde{i}}(\lambda),\, y_R^{\tilde{i}}(\lambda))}\, \omega
\label{croftongencov}
\end{equation}
The integral is taken over any smooth two-dimensional surface within the four-dimensional kinematic space with boundaries at the prescribed limits. Another way to characterize the region of integration is to say that we integrate over the image of any homotopy, which deforms $\{(y_L^{\tilde{i}}(\lambda),\, y_R^{\tilde{i}}(\lambda))\}_\lambda$ to $\{(y_L^o(\lambda),\, y_R^o(\lambda))\}_\lambda$ in kinematic space. 

To isolate the length of a given bulk curve, the limits of integration in (\ref{croftongencov}) should be set so that:
\begin{align*}
{\rm length}^{(o)} & = + \textrm{length of given curve} \\
{\rm length}^{(\,\tilde{i}\,)} & = - \textrm{length of given curve}
\end{align*}
The minus sign is easy to fix: the differential entropy formula incurs an extra minus sign when we switch the left and right endpoints of our oriented geodesics.

There are several ways to see this. A mechanical way is to view the bulk spacetime upside down: this switches the left and right endpoints of all intervals, but also switches the way in which we sweep the length of the curve, from clockwise to counterclockwise and vice versa. More formally, we can add to (\ref{diffent}) a total derivative term
\begin{align}
{\rm length} & = \phantom{-}
\int d\lambda\, \frac{dy_R^\mu}{d\lambda}\, \frac{\partial S(y_L(\lambda), y_R)}{\partial y_R^\mu} \Big|_{y_R = y_R(\lambda)} -
\int dS(y_L(\lambda), y_R(\lambda)) \nonumber \\
& = -
\int d\lambda\, \frac{dy_L^\mu}{d\lambda}\, \frac{\partial S(y_L, y_R(\lambda))}{\partial y_L^\mu} \Big|_{y_L = y_L(\lambda)}
\end{align}
and observe that the resulting integrand has the same form as in (\ref{diffent}), except for the switch $y_L(\lambda) \leftrightarrow y_R(\lambda)$ and the minus sign. We will shortly recognize the same fact from yet another perspective. 

It is useful to introduce a special notation for the reversal of endpoints:
\begin{equation}
\tilde{\phantom{.}}
\, : \quad (z_L, \bar{z}_L, z_R, \bar{z}_R) \, \longrightarrow \, (z_R, \bar{z}_R, z_L, \bar{z}_L)
\label{reversal}
\end{equation}
In eq.~(\ref{croftongencov}), we anticipated this notation as well as the following conclusion: the lower limit of integration should be a trajectory of geodesics that are NVA to the curve---i.e., which satisfy eq.~(\ref{diffent})---but with their endpoints reversed. 

In summary, the general covariant version of the Crofton formula reads:
\begin{equation}
{\rm length} 
= \frac{1}{2} \int^{(y_L^o(\lambda),\, y_R^o(\lambda))}_{(y_L^{\tilde{i}}(\lambda),\, y_R^{\tilde{i}}(\lambda))}\, \omega\,.
\label{croftoncov}
\end{equation}
Here $(y_L^o(\lambda), y_R^o(\lambda))$ and $(y_L^i(\lambda), y_R^i(\lambda))$ are any two smooth families of geodesics that are NVA to the bulk curve, $\tilde{\phantom{.}}$ is the endpoint reversal map defined in (\ref{reversal}) and the integral is carried out over any smooth two-dimensional submanifold of kinematic space with the prescribed boundaries, i.e. the image of a homotopy from $\{(y_L^{\tilde{i}}(\lambda),\, y_R^{\tilde{i}}(\lambda))\}_\lambda$ to $\{(y_L^o(\lambda),\, y_R^o(\lambda))\}_\lambda$.

\subsection{Comments}

Formula~(\ref{croftoncov}) merely rewrites eq.~(\ref{diffent}) using Stokes's theorem. We will see that it becomes much sharper when we apply it in pure AdS$_3$. Before that, however, we pause for a few comments about the application of (\ref{croftoncov}) in general asymptotically AdS$_3$ geometries:

\paragraph{A large freedom}
The length of a given bulk curve can be computed using formula~(\ref{croftoncov}) in multiple ways. First, we can choose any set of NVA geodesics on either limit of integration; both choices have a freedom described at the end of Sec.~\ref{NVACondition}. Further, we have a freedom of completing the domain of integration in any smooth way. In the static formula~(\ref{crofton}), all this freedom was killed off by restricting to quantities defined on a static slice.

\paragraph{Two branches of differential entropy, unified} We observed in Sec.~\ref{NVACondition} that for each curve there are two classes of differential entropy formulae, which involve geodesics that are tangent to the outgoing and ingoing orthogonal lightsheets. Eq.~(\ref{croftongen}) gives an opportunity to unify them: we can choose $(y_L^o(\lambda), y_R^o(\lambda))$---the NVA geodesics for the upper limit of integration---from the outgoing family and choose the endpoint-reversed lower limit $(y_L^i(\lambda), y_R^i(\lambda))$ from the ingoing family. The superscripts in the notation of eqs.~(\ref{croftongencov}, \ref{croftoncov}) anticipated this choice. At the level of eq.~(\ref{diffent}), the two families of NVA geodesics were not smoothly deformable into each other, being connected only through their joint special case of tangent geodesics. Going to the Crofton formula reveals that they form a boundary of a common smooth submanifold of kinematic space.

\paragraph{Four perspectives on the covariant Crofton formula}
Formula~(\ref{croftongen}) can be rewritten in other equivalent ways, which are generated by time reversal and parity. For clarity, we will apply these transformations passively, i.e. keeping the bulk curve fixed and changing perspective. Applied this way, time reversal $T$ simply swaps the ingoing and outgoing lightsheets: $i \leftrightarrow o$. Parity $P$, in turn, changes the sign of the line element along the curve, as well as swapping the left and right endpoints of all geodesics as in eq.~(\ref{reversal}). All in all, $T$ and $P$ generate these four Crofton formulae:
\begin{equation}
\stackrel{P}{\to}\,
\frac{1}{2} \int^{(y_L^o(\lambda),\, y_R^o(\lambda))}_{(y_L^{\tilde{i}}(\lambda),\, y_R^{\tilde{i}}(\lambda))}\!\!\omega 
\,\stackrel{T}{\to}\,
\frac{1}{2} \int^{(y_L^i(\lambda),\, y_R^i(\lambda))}_{(y_L^{\tilde{o}}(\lambda),\, y_R^{\tilde{o}}(\lambda))}\!\!\omega 
\,\stackrel{P}{\to}\,
- \frac{1}{2} \int^{(y_L^{\tilde{i}}(\lambda),\, y_R^{\tilde{i}}(\lambda))}_{(y_L^o(\lambda),\, y_R^o(\lambda))}\!\!\omega 
\,\stackrel{T}{\to}\,
- \frac{1}{2} \int^{(y_L^{\tilde{o}}(\lambda),\, y_R^{\tilde{o}}(\lambda))}_{(y_L^i(\lambda),\, y_R^i(\lambda))}\!\!\omega 
\,\stackrel{P}{\to}
\label{transformations}
\end{equation}
This transformation law under time reversal and parity is why we think it preferable to take the upper and lower integration limits in (\ref{croftoncov}) from distinct (ingoing and outgoing) NVA families. 

\section{Covariant Crofton formula in pure AdS$_3$}
\label{ads3sec}

Two dramatic simplifications occur in pure AdS$_3$: 

\paragraph{The first simplification} is that the entanglement entropy of an interval decomposes into a left-moving and a right-moving component (see e.g. \cite{cardycalabrese, opeblocks2}):
\begin{equation}
S(z_L, \bar{z}_L, z_R, \bar{z}_R) = s(z_L, z_R) + \bar{s}(\bar{z}_L, \bar{z}_R) \equiv
\frac{c}{6} \log\frac{\sin (z_R - z_L)/2}{\mu} + \frac{c}{6} \log\frac{\sin (\bar{z}_R - \bar{z}_L)/2}{\mu}
\,\,\,\,
\label{sdecomp}
\end{equation}
The decomposition is a consequence of the unbroken $SO(2,1) \times SO(2,1)$ global symmetry of the CFT$_2$. With the Brown-Henneaux relation \cite{brownhenneaux} $c = 3L_{\rm AdS}/2G_N$ and our convention $4G_N \equiv 1$, the coefficients in front of the logarithms are simply $L_{\rm AdS}$. In subsequent formulas for bulk lengths, we will not write down the explicit units of $L_{\rm AdS}$. Owing to eq.~(\ref{sdecomp}), the integrand of the Crofton formula becomes:
\begin{equation}
\omega_{{\rm AdS}_3} = (\partial_L \partial_R s) dz_L dz_R + (\bar{\partial}_L \bar{\partial}_R \bar{s}) d\bar{z}_L d\bar{z}_R
\label{omegaads3}
\end{equation}
Note that (\ref{omegaads3}) is invariant under four independent copies of circle reparameterizations, acting on each coordinate separately. 

\paragraph{The second simplification} concerns the freedom of choosing the integration domain for the Crofton formula. Both limits of integration in (\ref{croftoncov}) are subject to an ambiguity, which is parameterized by a single function on a circle. This function is the null rapidity parameter that separates the NVA geodesic from the tangent geodesic at a each point on the curve. It can be chosen freely everywhere along the curve, subject only to a continuity requirement. One may ask whether this large freedom stabilizes (leaves invariant) some bulk object or collection of objects, other than the given bulk curve itself. Identifying such a fixed set of the NVA freedom would allow us to unify all formulas~(\ref{croftoncov}) and organize them more meaningfully. As it turns out, the global symmetries of AdS$_3$ allow us to do just that.

\subsection{Integral over null planes}

Recall that all geodesics which are NVA to a given bulk curve at a point $\lambda$ are locally related to one another by null rotations. Because AdS$_3$ is a homogeneous space, nothing contaminates this statement further away from $\lambda$. In particular, all geodesics that are NVA to a given point on a curve are related by a null rotation \emph{globally}. We said previously that a null rotation preserves a null vector, just like rotations and boosts preserve timelike and spacelike vectors, respectively. In a homogeneous space like AdS$_3$, this means that a null rotation preserves a whole, globally defined plane orthogonal to its fixed null vector. Planes orthogonal to null vectors are called null planes; they are generated by one spacelike and one null vector.\footnote{Null planes are a null generalization of $\mathbb{H}_2$ and ${\rm AdS}_2$-hyperplanes of AdS$_3$, which are generated by two spacelike vectors (respectively one spacelike and one timelike vector). In standard embedding coordinates, a null plane is given by $N \cdot X \equiv (N_{-1}, N_0, N_1, N_2) \cdot (X_{-1}, X_0, X_1, X_2) = 0$ with normal vector $N \cdot N = 0$; this is in contrast to $\mathbb{H}_2$ and ${\rm AdS}_2$-hyperplanes described by the same planar equation with $N \cdot N \lessgtr 0$.} These null planes are the fixed sets of the NVA freedom, which we described abstractly in the previous paragraph. 

To understand this in greater detail, refer to Fig.~\ref{NullSheetsNVAGeodesics} and consider the NVA geodesics tangent to the outgoing lightsheet at $\lambda$. In the notation of the paragraph `Null rotations' (Sec.~\ref{NVACondition}), the tangent vectors of all such NVA geodesics are orthogonal to the null vector $n_o$ and, if they are properly normalized, they differ from one another only by multiples of $n_o$. In other words, the NVA geodesics span out a plane generated by the curve's tangent $t$ and by $n_o$---a plane orthogonal to the null normal vector $n_o$. All this is to say that the null plane is fixed under our NVA null rotation freedom even as individual geodesics contained in it transform into one another. For future use, we note that the null plane has exactly one lightray (the one generated by the normal vector $n_o$) in common with the bulk curve's outgoing lightsheet and that it is tangent to (shares a single point with) the bulk curve.

\begin{figure}[t!]
\centering
\includegraphics[width=.7\textwidth]{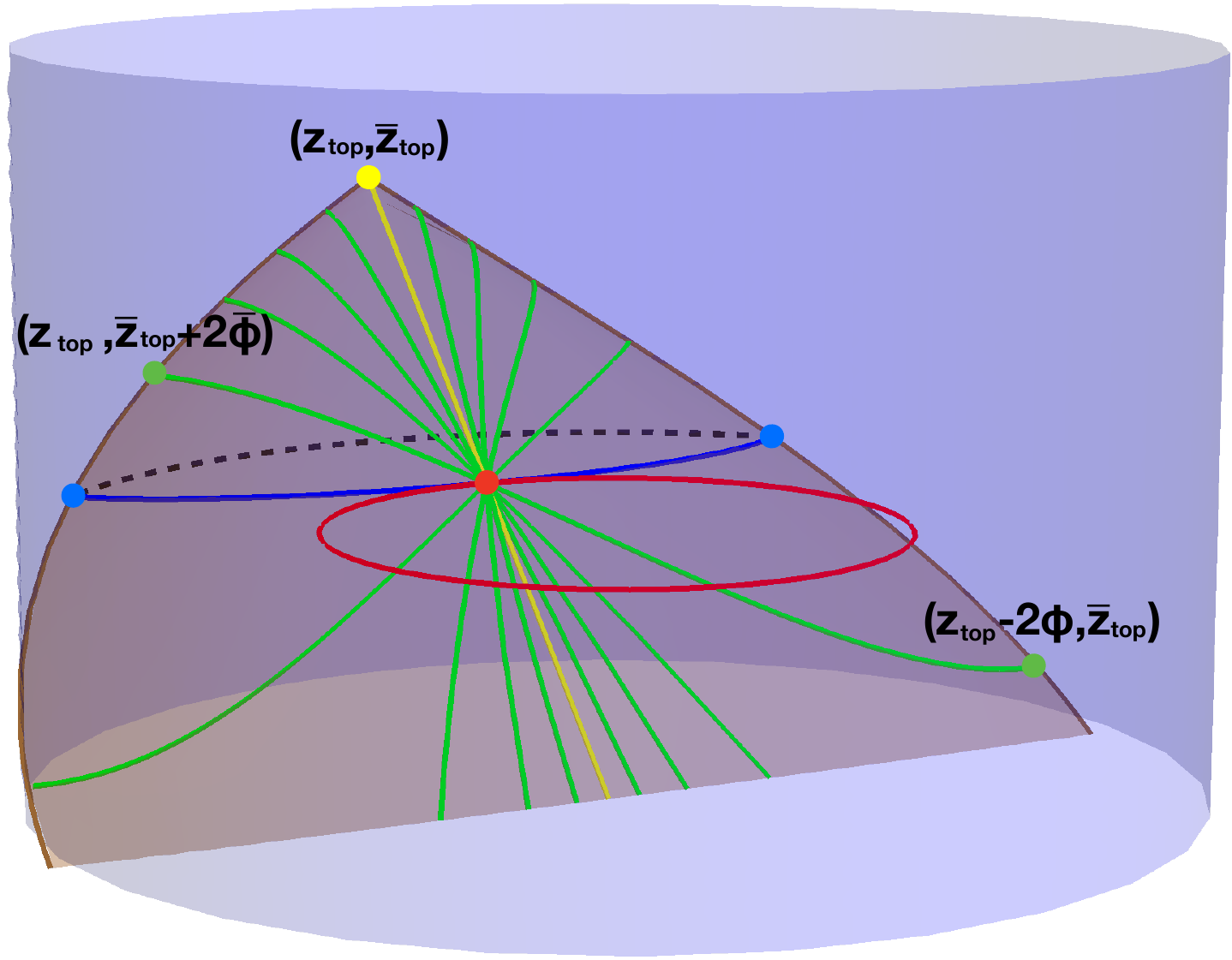}
\caption{The geodesics from the outgoing family, which are NVA to the bulk curve (red) at a common point. In pure AdS$_3$, such geodesics are contained in and span the null plane, which is generated by the orthogonal outgoing lightray $n_o$ (yellow) and the tangent geodesic (blue). All these NVA geodesics end on boundary lightrays $z_R = z_{\rm top}$ and $\bar{z}_L = \bar{z}_{\rm top}$, which is where the null plane meets the asymptotic boundary. The coordinates $(z_R, \bar{z}_L) = (z_{\rm top}, \bar{z}_{\rm top})$ are therefore common to all these NVA geodesics, independent of the gauge freedom parameterized by $\phi$ and $\bar{\phi}$. }
\label{GaugeInvariance}
\end{figure}

A null plane meets the asymptotic boundary of AdS$_3$ on two boundary null rays:
\begin{equation}
z = z_{\rm top} \qquad {\rm and} \qquad \bar{z} = \bar{z}_{\rm top}
\label{nullborders}
\end{equation}
We labeled the asymptotic borders of the null plane with the subscript `top' because these boundary null rays meet at the top of the null plane, which is a boundary point with coordinates $(z_{\rm top}, \bar{z}_{\rm top})$; see Fig.~\ref{GaugeInvariance}. The location of the top completely specifies the null plane, so $z_{\rm top}$ and $\bar{z}_{\rm top}$ are good coordinates on the space of all null planes in AdS$_3$.

Because the NVA geodesics tangent to the outgoing (respectively ingoing) lightsheet at $\lambda$ never leave the null plane, they must begin on one and end on the other of the two loci in (\ref{nullborders}). In particular, we must have either
\begin{align}
z_R = z_{\rm top} \qquad & {\rm and} \qquad \bar{z}_L = \bar{z}_{\rm top} \qquad {\rm or} \nonumber \\
z_L = z_{\rm top} \qquad & {\rm and} \qquad \bar{z}_R = \bar{z}_{\rm top}.
\end{align} 
A quick inspection reveals that the upper case applies to NVA geodesics tangent to the outgoing lightsheet while the lower case is valid for the ingoing family. Focusing on the upper (outgoing) case, observe that the extreme limit of exercising our null rotation freedom at $\lambda$ will produce a geodesic which is still NVA to the bulk curve but which becomes lightlike. This is the unique lightray common to the null plane and the bulk curve's orthogonal lightsheet: the lightray through $\lambda$ shot in the direction $n_o$. This null ray reaches the boundary precisely at $(z_{\rm top}, \bar{z}_{\rm top})$. In other words, the top of the null plane fixed by the NVA freedom is where the lightray generated by $n_o$ arrives at the asymptotic boundary. 

These observations identify a crisp common feature of all possible integration limits in~(\ref{croftoncov}) in CFT$_2$ vacuum / pure AdS$_3$. Whatever contour
\begin{equation}
(y_L^o(\lambda),\, y_R^o(\lambda)) = (z^o_L(\lambda), \bar{z}^o_L(\lambda), z^o_R(\lambda), \bar{z}^o_R(\lambda)) 
\label{outgoingtops}
\end{equation}
we choose for the upper limit of integration, we know that $(z^o_R(\lambda), \bar{z}^o_L(\lambda))$ must trace the boundary endpoints of orthogonal outgoing null rays shot out from the bulk curve. 
This is because the locus $(z^o_R(\lambda), \bar{z}^o_L(\lambda))$, interpreted as a family of tops of null planes tangent to the bulk curve, is invariant under the NVA freedom; see Fig.~\ref{GaugeInvariance}. 
Similarly, the lower limit of integration $(y_L^{\tilde{i}}(\lambda),\, y_R^{\tilde{i}}(\lambda))$ must be chosen so that its projection in kinematic space onto the $z_R, \bar{z}_L$ coordinates, 
\begin{equation}
(z_R^{\tilde{i}}(\lambda),\, \bar{z}_L^{\tilde{i}}(\lambda)) = (z_L^{i}(\lambda),\, \bar{z}_R^{i}(\lambda)),
\label{ingoingtops}
\end{equation}
traces the boundary endpoints of orthogonal ingoing null rays shot from the bulk curve. This is because (\ref{ingoingtops}) is the other family of tops of null planes tangent to the bulk curve---data that is, once again, unaffected by changes in the NVA null rapidity parameter. There are two continuous families of null planes tangent to the bulk curve---eq.~(\ref{outgoingtops}) and eq.~(\ref{ingoingtops})---because there are two null vectors orthogonal to the curve at each point.

\begin{figure}[t]
\centering
\includegraphics[width=.4\textwidth]{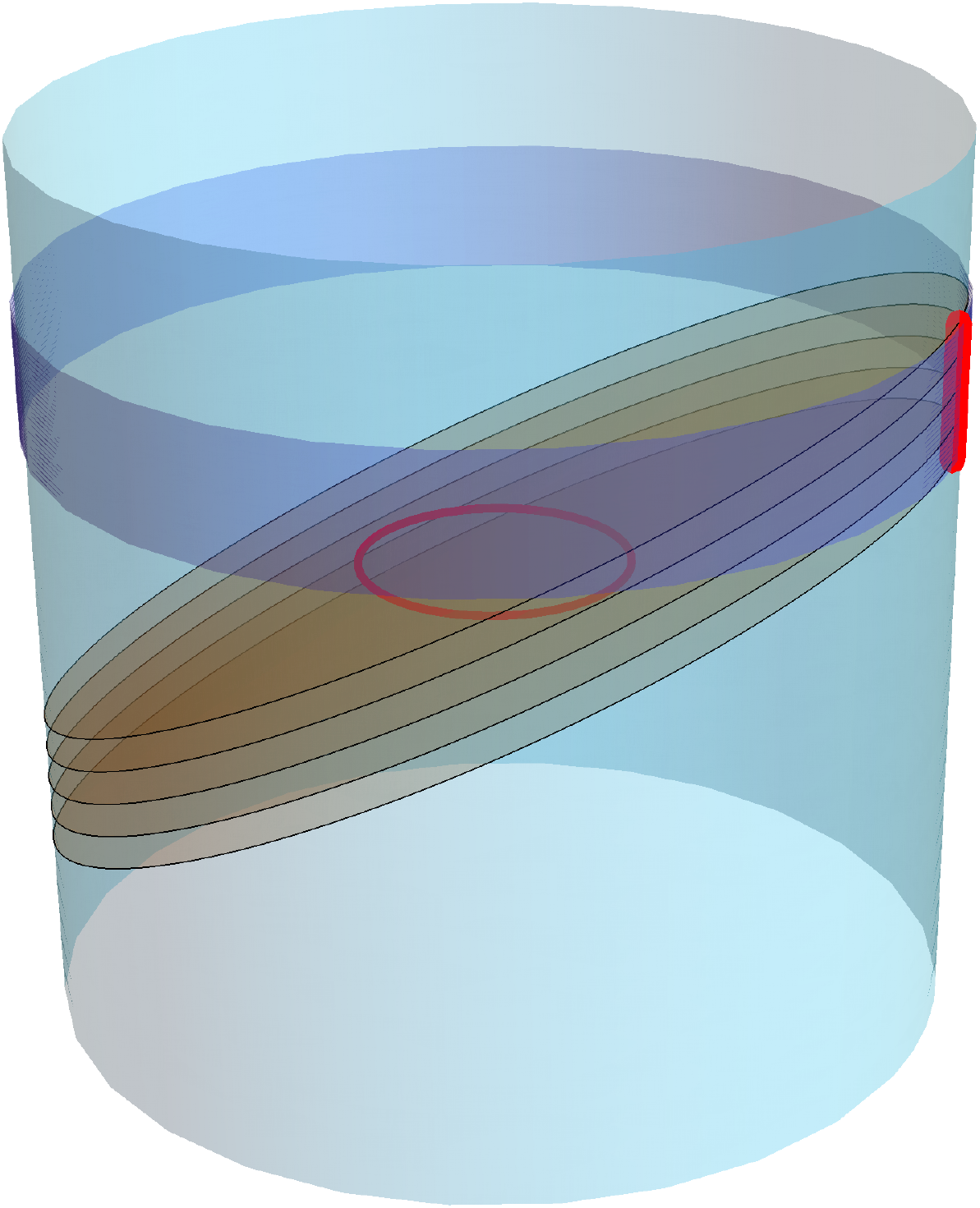}
\hfill
\includegraphics[width=.43\textwidth]{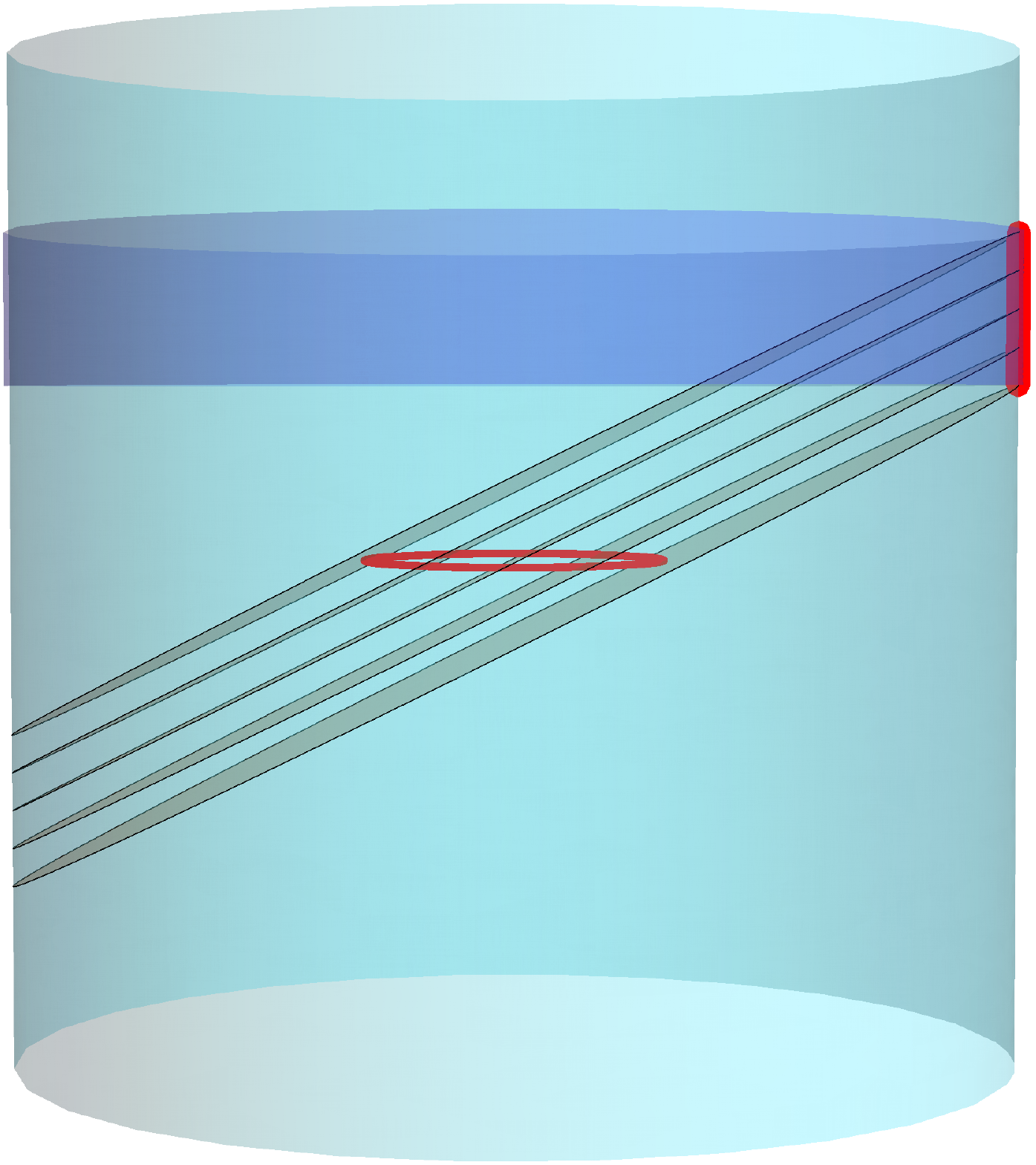}
\caption{The integration region in (\ref{croftonads3}) comprises null planes, which intersect the bulk curve. The tops of those null planes fall between the loci (\ref{outgoingtops}) and (\ref{ingoingtops}) and comprise points, which are neither entirely spacelike nor entirely timelike-separated from the whole curve.}
\label{intplanes} 
\end{figure}

The full domain of integration must have boundary points $(z_L, \bar{z}_R)$ fall in between those two curves. There are various ways to characterize this region; see Fig.~\ref{intplanes}. One is to say that it comprises those boundary points, which are spacelike-separated from at least one point on the curve and timelike-separated from at least one point on the curve:
\begin{equation}
\big(\cup_{p \in {\rm curve}} \textrm{causal future of $p$}\big) 
\cap
\big(\cup_{q \in {\rm curve}} \textrm{spacelike from $q$}\big)
\cap
\textrm{asymptotic boundary} 
\label{intregion1}
\end{equation}
It is possible to describe the region of integration by using the null cuts of \cite{nullcuts}. But the most succinct way is to observe that a null plane dropped from any point in (\ref{intregion1}) necessarily intersects the bulk curve. Indeed, the boundary of (\ref{intregion1}) are precisely those points, whose null planes barely skirt the curve. 

In summary, the Crofton formula (\ref{croftoncov}) in AdS$_3$ is an integral over null planes that intersect the bulk curve:
\begin{equation}
{\rm length} 
= \frac{1}{2} \int_\textrm{\{null planes that intersect the bulk curve\}} \, j^* \omega_{{\rm AdS}_3} 
\label{croftonads3}
\end{equation}
Here $j^*\omega_{{\rm AdS}_3}$ is the pullback of form~(\ref{omegaads3}) onto the two-dimensional space of null planes parameterized by $z_L$ and $\bar{z}_R$. We discuss this pullback in the next subsection.

\paragraph{Non-convex curves} 
In the static case, a generalization of formula~(\ref{crofton}) that applies to non-convex curves reads \cite{intgeom}:
\begin{equation}
{\rm length} 
= \frac{1}{4} \int_\textrm{\{geodesics that intersect the bulk curve\}} \, d\theta_R d\theta_L\, 
\frac{\partial^2 S(\theta_L, \theta_R)}{\partial \theta_R\, \partial \theta_L}\, n(\theta_L, \theta_R)
\label{croftonnc}
\end{equation}
Here $n(\theta_L, \theta_R)$ is the number of intersections of geodesic $(\theta_L, \theta_R)$ with the bulk curve on the static slice. For convex curves, this number is either 2 (for intersecting geodesics) or 0 (for non-intersecting ones), except for the codimension-1 set of tangent geodesics. Substituting these values of $n(\theta_L, \theta_R)$ recovers equation~(\ref{crofton}). 

It is easy to see that the generalization of (\ref{croftonads3}) to non-convex curves follows the same pattern:
\begin{equation}
{\rm length} 
= \frac{1}{4} \int_\textrm{\{null planes $\sigma$\}} \, j^* \omega_{{\rm AdS}_3}\, n(\sigma)\,,
\label{croftonads3nc}
\end{equation}
where $n(\sigma)$ is the number of intersections of the bulk curve with the null plane $\sigma$. For convex curves this number is again either 2 or 0, except for the codimension-1 set of null planes that are tangent to the bulk curve; this establishes the consistency of (\ref{croftonads3nc}) with (\ref{croftonads3}). For non-convex curves, one can prove eq.~(\ref{croftonads3nc}) by adding and subtracting to the curve geodesic segments that complete its convex cover. 

\paragraph{Remark}
Ref.~\cite{solanes} provides a general formula for the volume of an $m$-dimensional locus in $n$-dimensional hyperbolic space in terms of its intersections with $r$-dimensional hyperplanes, for any $r + m \geq n$. Our Crofton formula~(\ref{croftonads3nc}) is one natural generalization of that result to the Lorentzian context. It is interesting that the Lorentzian version singles out null planes as the homogeneous objects to be integrated. It should be straightforward to generalize~(\ref{croftonads3nc}) to higher-dimensional pure anti-de Sitter spaces, but we do not pursue it in this paper.

\subsection{Induced measure over null planes}

In eq.~(\ref{croftonads3}), we are instructed to integrate the pullback of $\omega$ onto the space of null planes. By what map 
\begin{equation}
j: \,\,\,\{ \textrm{null planes} \} \,\,\, \longrightarrow \,\,\,\textrm{kinematic space}
\end{equation}
are we pulling $\omega$ back?

On the boundary of the integration region, that is for null planes tangent to the bulk curve, map $j$ assigns to a null plane with top point $(z_R(\lambda), \bar{z}_L(\lambda))$ a geodesic that is NVA to the bulk curve at $\lambda$. Of course, if $(z^o_R(\lambda), \bar{z}^o_L(\lambda))$ is the boundary endpoint of the \emph{outgoing} orthogonal null ray then the NVA geodesic $j\big((z^o_R(\lambda), \bar{z}^o_L(\lambda))\big)$ must be tangent to the \emph{outgoing} lightsheet; an analogous consistency condition applies to the ingoing family. In the interior of the integration region, the assignment of geodesics to null planes is arbitrary except for a smoothness requirement. The `boundary conditions' for the embedding map $j$ are summarized by the following equations:
\begin{align}
j\big((z^o_R(\lambda), \bar{z}^o_L(\lambda))\big) & = (y_L^o(\lambda),\, y_R^o(\lambda)) 
\label{bc1} \\
j\big((z^{\tilde{i}}_R(\lambda), \bar{z}^{\tilde{i}}_L(\lambda))\big) & = (y_L^{\tilde{i}}(\lambda),\, y_R^{\tilde{i}}(\lambda)) = (y_R^i(\lambda),\, y_L^i(\lambda))
\label{bc2}
\end{align}
Naturally, different bulk curves will give rise to different embeddings $j$. 

To be more explicit, let us change coordinates from $z_L, \bar{z}_L, z_R, \bar{z}_R$ to $z_L, \bar{z}_R$ and:
\begin{equation}
\phi = (z_R - z_L)/2 \qquad {\rm and} \qquad \bar{\phi} = (\bar{z}_R - \bar{z}_L)/2.
\end{equation}
The two-dimensional surface in kinematic space, over which we carry out the integral in eq.~(\ref{croftoncov}), is now specified by two functions $\phi(z_R, \bar{z}_L)$ and $\bar{\phi}(z_R, \bar{z}_L)$. In terms of these functions, the embedding of the space of null planes in kinematic space is:
\begin{equation}
j(z_R, \bar{z}_L) = 
(z_L=z_R-2\phi(z_R, \bar{z}_L), \bar{z}_L, 
 z_R, \bar{z}_R=\bar{z}_L+2\bar{\phi}(z_R, \bar{z}_L)). 
\end{equation}
Pulling back $\omega_{{\rm AdS}_3} $ by this map, we obtain the integrand of the Crofton formula~(\ref{croftonads3nc}):
\begin{equation}
\label{CroftonIntegrand}
j^* \omega_{{\rm AdS}_3} = \frac{1}{2}\left(\partial_R\cot\bar{\phi}-\bar{\partial}_L\cot\phi\right) dz_R d\bar{z}_L
\end{equation}
We refer to the functions $\phi(z_R, \bar{z}_L)$ and $\bar{\phi}(z_R, \bar{z}_L)$ collectively as a `gauge freedom.'

Conditions~(\ref{bc1}, \ref{bc2}) say that on the boundary of the integration region, i.e. for null planes tangent to the bulk curve, the parameters $\phi$ and $\bar{\phi}$ are not independent. Their relation serves to impose the NVA condition. For a geodesic that already lives on a null plane tangent to the bulk curve, all that remains to satisfy the NVA condition is to insure that the geodesic passes through the curve. Therefore, equation~(\ref{bc1}) simply states that the geodesic $(y_L^o(\lambda),\, y_R^o(\lambda))$ meets the bulk curve at $\lambda$; see Fig.~\ref{GaugeInvariance}.

\subsection{Example}
\label{sec:example}

We exemplify the above results by computing the circumference of a circle in AdS$_3$ in arbitrary gauge. Let us use coordinates
\begin{equation}
\label{GlobalCoordinates}
ds^2 = -(1+R^2)\, dt^2 + (1+R^2)^{-1} dR^2 + R^2\, d\theta^2,
\end{equation}
remembering that dimensions of length are supplied by factors of $L_{\rm AdS}$. 
The null rays orthogonal to the circle of radius $R = R_0$ at $t = 0$ reach the asymptotic boundary at $t = \cot^{-1} R_0$ and $t = \pi - \cot^{-1} R_0$. Therefore, the region of integration in eq.~(\ref{croftonads3}) will cover:
\begin{equation}
\mathcal{R}: \quad \cot^{-1}R_0 \leq t_{\rm top} = \frac{z_R-\bar{z}_L}{2} \leq \pi-\cot^{-1}R_0 
\quad {\rm and} \quad 
0 \leq \theta_{\rm top} = \frac{z_R+\bar{z}_L}{2} \leq 2\pi.
\end{equation}
The choice of gauge $\phi(z_R, \bar{z}_L)$ and $\bar{\phi}(z_R, \bar{z}_L)$ is arbitrary in the interior of the integration region, but on the boundary we must ensure that the selected geodesic (choice of gauge on the null plane) touches the circle. This requirement becomes:
\begin{equation}
\label{NVAC}
\cot\phi + \cot \bar{\phi} = \pm 2 R_0\,,
\end{equation}
where the upper sign holds for the outgoing family and the lower sign for the ingoing family of null planes. Substituting all these into (\ref{croftonads3}) gives:
\begin{align}
&
\frac{1}{2} \int_{\mathcal{R}} \frac{dz_R d\bar{z}_L}{2} \left(\partial_R\cot\bar{\phi}-\bar{\partial}_L\cot\phi\right) 
\nonumber \\
= & \, -\frac{1}{4}\int_{\mathcal{R}}d\theta_{\rm top} dt_{\rm top} \left( \partial_{\theta_{\rm top}}(\cot\bar{\phi}-\cot{\phi})+\partial_{t_{\rm top}}(\cot\phi+\cot\bar{\phi}) \right)  
\nonumber \\
= & \,
-
\frac{1}{4} \int_{\{t_{\rm top} = \pi - \cot^{-1}R_0\}} d\theta_{\rm top} 
\Big(\! \cot\bar{\phi} + \cot\phi\Big)
+ \frac{1}{4} \int_{\{t_{\rm top} = \cot^{-1}R_0\}} d\theta_{\rm top} 
\Big(\! \cot\bar{\phi} + \cot\phi\Big)
\nonumber \\
= & \, 2\pi R_0
\label{examplecomp}
\end{align}

\section{Discussion}

We have obtained Crofton formulas (\ref{croftoncov}), which compute lengths of spacelike curves in horizonless but otherwise general, asymptotically AdS$_3$ geometries. Crofton formulas are integrals over geodesics, which satisfy a certain relation to the curve. In (\ref{croftoncov}), the requisite relation is that the geodesics are part of a homotopy, which deforms one loop in the space of geodesics into another. The beginning (ending) loop in this homotopy consists of geodesics, which are null vector-aligned (NVA) to the ingoing (outgoing) orthogonal lightsheet of the curve. As the geodesics we consider are oriented, we must in addition stipulate that in one of the two loops the geodesics are endpoint-reversed. 

We took on the problem of covariantizing the Crofton formula in the hope of informing a future quest for an understanding of bulk \emph{time}, akin to the present understanding of how holographic bulk \emph{space} emerges from quantum entanglement in the boundary theory. Let us list the lessons we reaped from this exploration:
\begin{itemize} 
\item Underlying the static differential entropy formula (\ref{diffentstatic}) is the notion of tangency between a curve and a geodesic segment. In the covariant case, the tangency condition is replaced by a weaker one: that the \emph{orthogonal lightsheets} shot from the curve and from the geodesic be tangent. This is the null vector alignment (NVA) condition \cite{hholes}. Thus, in going from the static (\ref{diffentstatic}) to the covariant (\ref{diffent}) differential entropy, we effectively trade geodesic segments for local patches of lightsheets.
\item Weakening the required notion of tangency provides a large freedom in the differential entropy formula and an even larger `gauge freedom' for Crofton formulas. Geometrically, this freedom is generated by null rotations in the bulk---the local symmetry that stabilizes a local piece of a lightsheet.
\item In the interior of the integration region in the Crofton formula, the `gauge freedom' means that we no longer integrate over intersecting geodesics, but over a more abstractly defined collection of them. The integral covers the image of a homotopy: a continuous way of deforming geodesics which are NVA to the curve's ingoing lightsheet into those, which are NVA to the curve's outgoing lightsheet.
\item Contrary to what one may have inferred from the static Crofton formula~(\ref{crofton}), even in the static case there is nothing special about geodesics which intersect the curve.  The example discussed in Sec.~\ref{sec:example} is a case in point: any asymmetric choice of `gauge' $\phi \neq \bar{\phi}$ will bring into the integral (\ref{examplecomp}) geodesics that do not intersect the static circle.  
\end{itemize}

Characterizing the integration region as a homotopy that links the two sets of NVA geodesics is not very revealing because it follows so directly from applying Stokes's theorem to the differential entropy formula (\ref{diffent}). For general geometries, we have not found a crisper characterization of the integration region. If it can be formulated, it must rely on the NVA condition which, as we explained in Sec.~\ref{NVACondition}, is a manifestation of the null rotation symmetry of sufficiently small neighborhoods of points on the curve. This suggests that a more satisfactory reading of formula~(\ref{croftoncov}) will rely on a deeper holographic understanding of bulk null rotations and of their fixed axes---bulk null rays. We believe that a search for a conceptual, perhaps information theoretic, boundary understanding of bulk null rays is a promising direction for future research. 

In pure AdS$_3$, however, our results simplify dramatically:
\begin{itemize}
\item Instead of local patches of lightsheets, in pure AdS$_3$ we may work with \emph{globally} defined lightsheets (null planes), which are fixed sets of null rotations. 
\item Null planes give a natural way of parameterizing the two-dimensional integral (\ref{croftoncov}). No matter how we exploit the `gauge freedom' in eq.~(\ref{croftoncov}), the integral always covers the same null planes.
\item The final answer, eq.~(\ref{croftonads3}) with measure (\ref{CroftonIntegrand}), splits up into two separate pieces, which depend only on the left-moving (respectively right-moving) component of the `gauge choice,' $\phi$ (respectively $\bar\phi$). This is a consequence of the unbroken $SO(2,1) \times SO(2,1)$ global conformal symmetry. 
\item The two summands in (\ref{croftonads3}) are only `coupled' at the boundary of the integration region where the NVA condition is imposed, as in eq.~(\ref{NVAC}) in the example in Sec.~\ref{sec:example}. 
\end{itemize}
These simplifications occur because a null rotation in pure AdS$_3$ stabilizes a globally defined null plane. This is why in pure AdS$_3$ null planes become the basic objects that label the geodesics to be integrated over. In our view, this fact contains some hint for covariantizing the program of deriving geometry from quantum entanglement. For example, could one devise tensor networks, whose individual tensors correspond to null planes instead of points \cite{briantns, errortn, pluperfect} or geodesics \cite{tnks} on a spatial slice?

We close with two further remarks concerning the Crofton formulas~(\ref{croftonads3}) for pure AdS$_3$. First, the formulas apply equally well in Ba{\~n}ados geometries \cite{banados}, i.e. locally AdS$_3$ spacetimes obtained from pure AdS$_3$ by large diffeomorphisms. This is because, in a passive reading, a large diffeomorphism does not affect a null plane, so the integral~(\ref{croftonads3}) is taken over the same region of integration. The only thing that does change is the boundary parameterization of the null planes and geodesics. Indeed, Ba{\~n}ados geometries are holographic duals of Virasoro descendants of the CFT$_2$ ground state, which are related to the ground state by a finite conformal transformation that sends:
\begin{equation}
z \to f(z) \qquad {\rm and} \qquad \bar{z} \to \bar{f}(\bar{z}).
\end{equation}
As we remarked in Sec.~\ref{4dksintro}, the measure~(\ref{4dcroftonform}) is invariant under such a reparameterization. 

Second, the two components of the pulled-back measure~(\ref{CroftonIntegrand}) can be considered independently. As explained in Refs.~\cite{mbc, sewingkit}, the two-form (\ref{omegaads3}) can be identified with the modular Berry curvature and, as a consequence, its integrals compute modular Berry transformations---generalizations of the familiar Berry phases \cite{berry, wilczekzee}, which are induced by varying the \emph{modular} Hamiltonians. For example, equation~(\ref{croftongen}) computes the difference between the modular Berry transformations induced by drawing $\theta_R$-dependent modular Hamiltonians from the continuous family of intervals $\{(\theta_L^{\rm upper}(\theta_R), \theta_R)\}|_{\theta_R}$, relative to the family of intervals $\{(\theta_L^{\rm lower}(\theta_R), \theta_R)\}|_{\theta_R}$. Eq.~(\ref{croftongen}) evaluates to a difference of two lengths because in this case the modular Berry transformation is a translation along the geodesic. 

More generally, a modular Berry transformation lives in the commutant of the given modular Hamiltonian. In pure AdS$_3$, global conformal symmetry $SO(2,2)$ alone guarantees that this commutant must be at least as large as $SO(1,1) \times SO(1,1)$. Translations along the geodesic correspond to one combination of the two $SO(1,1)$s; the other one corresponds to modular boosts generated by the modular Hamiltonian itself. The way to capture this effect is to flip the relative sign in eq.~(\ref{omegaads3}) or in (\ref{CroftonIntegrand}). Thus, integrals of the form 
\begin{equation}
\omega^{(-)}_{{\rm AdS}_3} = - (\partial_L \partial_R s) dz_L dz_R + (\bar{\partial}_L \bar{\partial}_R \bar{s}) d\bar{z}_L d\bar{z}_R
\label{omegaminus}
\end{equation}
or 
\begin{equation}
\label{CroftonIntegrandminus}
j^* \omega^{(-)}_{{\rm AdS}_3} = \frac{1}{2}\left(\partial_R\cot\bar{\phi}+\bar{\partial}_L\cot\phi\right) dz_R d\bar{z}_L
\end{equation}
compute the component of a modular Berry transformation, which is an evolution with the modular Hamiltonian over some finite amount of modular time. In the bulk of AdS$_3$, this is a finite boost in the plane orthogonal to the geodesic. In the example of Sec.~\ref{sec:example}, we can compute this modular boost by substituting the integrand of eq.~(\ref{examplecomp}) with (\ref{CroftonIntegrandminus}) and setting the limits of integration to any generic family of geodesics that are NVA to the circle so long as $\phi \neq \bar{\phi}$. More details on the bulk picture of this construction will be given in \cite{zizhi}.


\section*{Acknowledgments}
We are grateful to Vijay Balasubramanian, Bowen Chen, Alberto G{\"u}ijosa, Moshe Rozali and Gongwang Yan for discussions. 
BC acknowledges the hospitality of MITP during the `Modern Techniques for AdS and CFT' program, of the Galileo Galilei Institute during the `Entanglement in Quantum Systems' program, of TSIMF Sanya during the `Workshop on Black Holes and Holography,' of AEI Postdam during the `Tensor Networks: From Simulations to Holography II' program, of Fudan University during the `String Theory and Quantum Field Theory' program, as well as the hospitality of the Institute for Advanced Study (Princeton), the University of Pennsylvania, MIT, and the Weizmann Institute, where part of this work was completed. BC is supported by the Startup Fund from Tsinghua University and by the Young Thousand Talents Program. The work of YDO was partially supported by Mexico's National Council of Science and Technology (CONACyT) Grant 238734 and DGAPA-UNAM Grant IN107115.


\end{document}